\documentclass{aastex63}

\received{2020 October 22}
\revised{20xx}
\accepted{20xx} 
\submitjournal{ApJS}   
\shorttitle{Draft}
\shortauthors{Sofue et al.} 

\begin{document}

\title{Atlas of CO-Line Shells and Cavities around Galactic Supernova Remnants with FUGIN\footnote{FUGIN: FOREST (FOur beam
REceiver System on the 45-m Telescope) Unbiased Galactic plane Imaging with the Nobeyama 45-m telescope}} 

\correspondingauthor{Yoshiaki Sofue}
\email{sofue@ioa.s.u-tokyo.ac.jp}
\author[0000-0002-4268-6499]{Yoshiaki Sofue} 
\affiliation{Institute of Astronomy, The University of Tokyo, 2-21-1 Mitaka, Tokyo 181-8588, Japan }
\author[0000-0003-1487-5417]{Mikito Kohno} 
\affiliation{Astronomy Section, Nagoya City Science Museum, 2-17-1 Sakae, Naka-ku, Nagoya, Aichi 460-0008, Japan} 
\author[0000-0001-8964-2066]{Tomofumi Umemoto} 
\affiliation{Nobeyama Radio Observatory, National Astronomical Observatory of Japan (NAOJ), National Institutes of Natural Sciences (NINS), 462-2 Nobeyama, Minamimaki, Minamisaku, Nagano 384-1305, Japan}
\affiliation{Department of Astronomical Science, School of Physical Science, SOKENDAI (The Graduate University for Advanced Studies), 2-21-1 Osawa, Mitaka, Tokyo 181-8588, Japan}


\def\vlsr{v_{\rm LSR}} \def\Msun{M_\odot} \def\vr{v_{\rm r}}
\def\deg{^\circ}  \def\Vrot{V_{\rm rot}}  
\def\co{$^{12}$CO }  \def\coj{$^{12}(J=1-0)$ }
\def\coth{$^{13}$CO } \def\coei{C$^{18}$O }
\def\Xco{X_{\rm CO}}   \def\Tb{T_{\rm B}}
\def\Htwo{H$_2$ }   \def\Kkms{K km s$^{-1}}  
\def\Hcc{{\rm H \cm^{-3}}} 
\def\Msun{M_\odot}  
 \def\kms{km s$^{-1}$} \def\Ico{I_{\rm CO}} 
\def\mH{m_{\rm H}}  \def\Htwo{H$_2$ }    
\def\Ico{I_{\rm CO}} 
\def\htwo{H$_2$} 
\def\vlsr{v_{\rm lsr}}
\def\Tb{T_{\rm B}}  
 \def\mH{m_{\rm H}}   
 \def\red{}
   
\begin{abstract} 
\red{A morphological} search for molecular shells and cavities was performed around 63 Galactic supernova remnants (SNR) at $10\deg \le l \le 50\deg$, $|b|\le 1\deg $using the FUGIN (FOREST Unbiased Galactic Imaging survey with the Nobeyama 45-m telescope) CO line data at high angular ($20$\arcsec) and velocity (1.3 \kms) resolutions. 
The results are presented as supplementary data for general purpose for investigations of the interaction between SNRs and interstellar matter in the form of an atlas of CO-line maps superposed on radio continuum maps at 20 cm along with a list of their kinematic distances determined from CO-line radial velocities.
({\bf Full atlas including all figures is available in:
 https://nro-fugin.github.io/2020-apjs-CO-Shell-Atlas-SNR-FUGIN-IX.pdf})

\end{abstract}    

\section{Introduction}
 
\red{The interaction} between shock waves of supernova remnants (SNR) and molecular clouds (MC) has been a long-standing issue in the physics of the interstellar medium (ISM) \citep{chev1977ARA&A..15..175C,chev1999ApJ...511..798C,shull1980,lucas+2020}.
The major concerns about the interaction are the generation of interstellar turbulence
\citep{kilpa2016ApJ...816....1K},
triggering or suppression of star formation 
\citep{mckee1977ApJ...218..148M,cox1999ApJ...524..179C,seta2004AJ....127.1098S},
and cosmic ray acceleration
\citep{fujita2009ApJ...707L.179F,2018ApJ...864..161K,2019ApJ...885..129M,sano2019ApJ...876...37S}.

Extensive observations of association of molecular clouds with well studied SNRs have been obtained in the last decades by molecular line observations 
\citep{tate1990A&A...237..189T,koo1997ApJ...485..263K,tian2007A&A...474..541T,rana2018MNRAS.477.2243R,lee+2020arXiv201107711L}. 
The current 'association' has been discussed mainly by the coincidence of the distance of an SNR measured by some means with the kinematic distance of the cloud from radial velocity, leaving large uncertainty on the order of $\sim 1$ kpc.
On the other hand, the association based on the morphological shell structure concentric to the SNR's shock front has been obtained in few cases.

In this paper, we perform a systematic search for CO-line shells and/or cavities based on morphological association with the SNRs listed in the Green's catalogue ({http://www.mrao.cam.ac.uk/surveys/snrs}) 
\citep{green2009BASI...37...45G,green1992MNRAS.254..686G,2019JApA...40...36G}.
We use \co and \coth ($J=1-0$) line channel maps from the FUGIN data set
\citep{2016SPIE.9914E..1ZM,2017PASJ...69...78U}.

The purpose of this paper is to present the result in the form of an atlas of the identified molecular cavities and shells, and to provide with a finding chart for general purpose of the research of the interaction between \red{SNRs and the ISM} in the Galactic disc. 

\section{Data}

\begin{table*}[htbp] 
\caption{Parameters of data sets}
\begin{flushleft}
\begin{tabular}{cccccccccccc}
\hline
Telescope/Survey & Line/Band &Effective   &  Velocity & References \\
& & Resolution & Resolution  &\\
\hline
\hline
Nobeyama 45-m/FUGIN &$^{12}$CO $J=$ 1--0  & 20\arcsec  & 1.3 $\>$km s$^{-1}$   & 
\citet{2017PASJ...69...78U}$^1$\\
&$^{13}$CO $J=$ 1--0 & 21\arcsec  &  1.3 $\>$km s$^{-1}$  &  
\citet{2017PASJ...69...78U} \\
\hline
VLA/VGPS & 21 cm   & \red{$\sim 1$\arcmin} &---&\cite{stil2006AJ....132.1158S}$^2$  \\
VLA/MAGPIS & 20 cm & $\sim 6$\arcsec & ---&\citet{2006AJ....131.2525H}$^3$  \\
Effelsberg 100-m/Galactic Plane & 21 cm  &$\sim 9.4$\arcmin 
&--- & \citet{1997AAS..126..413R}$^4$\\
\hline
\hline
\end{tabular}
\end{flushleft}
$[1]$ {http://nro-fugin.github.io}\\
$[2]$ {http://www.ras.ucalgary.ca/VGPS/VGPS\_data.html}\\
$[3]$ {https://third.ucllnl.org/gps/index.html}\\
$[4]$ {http://www3.mpifr-bonn.mpg.de/survey.html}\\
\label{obs_param}
\end{table*}

Table \ref{tab1} lists the SNRs from the Green's catalogue located in the FUGIN survey area at $10\deg \le l \le 50\deg$ and $|b|\le 1\deg$ and $190\deg\le l \le 230\deg$.
Figure \ref{GreenOnFugin} shows the positions of the SNRs on the color-coded maps of the peak $\Tb$ of the \co, \coth, and \coei line emission in the first Galactic quadrant \citep{2017PASJ...69...78U}.
In order to compare the distributions of the CO line emission with radio distribution of the SNRs, we extracted 21 cm radio continuum maps of the SNRs from the archival web sites of the Multi-Array Galactic Plane Imaging Survey (MAGPIS: \cite{2006AJ....131.2525H}), VLA Galactic Plane Survey (VGPS: \cite{stil2006AJ....132.1158S}), and 
the Effelsberg radio continuum survey \citep{1997AAS..126..413R}.
We summarize the parameters of data sets in Table \ref{obs_param}.

The FUGIN project provided with high-sensitivity, high-spatial and velocity resolution,  wide velocity ($482\ {\rm channels}\times 0.65$ \kms) and wide field ($40\deg \times 2\deg$ along the Galactic plane from $l=10\deg$ to $50\deg$) coverage by $(l,b,\vlsr:\Tb)$ cubes in the \co, \coth and \coei ($J=1-0$) lines. 
The full beam width at half maximum of the telescope was 15\arcsec at the \co ($J=1-0$)-line frequency, and the velocity resolution was 1.3 \kms. 
The effective beam size of the final data cube was 20\arcsec, and the rms noise levels were $\sim 1$ K.  
The final 3D FITS cube had a voxel size of $(\Delta l, \Delta b, \Delta \vlsr) =$ (8.5\arcsec, 8.5\arcsec, 0.65 \kms), which are available as the archival data.

 \section{Atlas} 

\subsection{\red{Identification}} 

We present the atlas of molecular cavities, shells and partial arcs apparently surrounding the SNRs by superposition of \co channel maps on radio continuum maps at 20 cm.  

The search for a CO shell associated with a SNR was done by the following procedure. Since the distance of a SNR is unknown, or uncertain even if it exists, so that its radial velocity is not known, the search for the shell structure was done in all the 462 channels of the CO data cube from -100 to 200 \kms by one channel after another for each SNR. 

First we display the radio continuum image on the screen, and superpose a channel map ($\Tb$ map) on the same screen. Then, the CO channel is changed from 1st to 462nd step by step. Numerous CO clouds and filaments will pass by, mostly fore- and background emissions, but at a certain velocity channel, a possible shell/cavity/arc appears apparently associated with the SNR's shell edge. 

Once such a candidate was found, its nearby channels are inspected more carefully, and the clearest shell feature was chosen as the associated shell, and its channel velocity was adopted as the radial velocity of the shell. 
This was repeated in \co and \coth cubes, each 462 channels, for all the 63 SNRs. 
The \coei data were not used for their too low brightness for the present purpose.  

Table \ref{tab1} in Appendix B lists the Galactic positions ($l,b)$, radial velocities ($\vlsr$), kinematic distances \red{($d$)}, and linear diameters ($D$) of the candidates cavities and/or shells of the analyzed SNRs.
\red{References to molecular-line observations, which report the association of the same or close radial velocities are cited in the last column. 
Objects without references are mostly new measurements currently with no information about molecular gas association.}

The measured results are presented in the form of $\Tb$ maps (sometimes $\Ico$ maps) in the \co line emission of the CO shells, arcs, and/or concentric alignment of clumps, as superposed on 20-cm radio continuum maps,
\red{in figures \ref{G1100} to 51 of Appendix C.}
We also present superposed $^{13}$CO and $^{12}$CO maps by R (red) and G (green) color-coded images in order to show the degree of condensation of the molecular gas density.  
 
\red{The association of the SNR and the identified molecular shell/cavity has been obtained purely by morphological inspection into the CO-line channel maps.
This means that, despite of coincidence of the concave edge of a CO cloud with the SNR's outer edge, the physical (true) association cannot be proved from the present analysis, which
applies to all the SNRs studied in this paper.
A direct way to prove the association is their distance coincidence on the line of sight with a sufficiently high accuracy, e.g. within an error of a few tens of pc.
However, the kinematical distance from the molecular line includes uncertainty of $\sim 1$ kpc; 
distance estimation by SNR's brightness to diameter ($\Sigma-D$) relation yields even larger uncertainty; 
and the method to measure the radio continuum absorption by an HI cloud includes a problem of the association of the HI and CO clouds themselves.}

\red{Thus, the atlas presents only the candidate molecular structures.
More advanced discussion about the interaction may be obtained by further, sophisticated observations of a signature of the physical compression by a shock wave such as shock-induced molecular lines \citep{koo1997ApJ...485..263K,ziurys1989ApJ...341..857Z,seta2004AJ....127.1098S,2013ApJ...774...10S}.}

\red{Nevertheless, it may be worthwhile to comment that the here determined radial velocities, hence kinematical distances, of the identified molecular shells for some typical SNRs are in good agreement with those currently reported in the literature such as G11.17-0.35 at $\vlsr = 33$ \kms \citep{kilpa2016ApJ...816....1K},
and W44 at $40-50$ \kms \citep{seta2004AJ....127.1098S}, for example. 
In table \ref{tab1} we cite more references, in which the same or close velocity clouds are identified by independent CO line observations.}

\subsection{Morphological classification}

\red{A SNR} interacting with a molecular cloud will deform the cloud to make a concave boundary with respect to the SNR center.
Thereby, the resulting cloud morphology will depend on the extent and density of the cloud. 
We categorize the structure of a shell or a cavity of the CO brightness distribution apparently surrounding a SNR as follows:

i) Cavity (Ca $\kappa$): When the cloud is extended more largely than the SNR size or comparable, a round cavity is created around the SNR due to dissociation of molecular gas and accumulation at the shock front. 
If the cloud size is sufficiently large, the cavity will be fully embedded in the cloud, making a round shape on the sky. 
We define such a case a cavity with completeness of 1 or 100\%, and introduce a completeness parameter or the shell measure, $\kappa=1$. 
If the cloud size is comparable or smaller, the dissociation and/or compression will take place  partially, forming open cavity to the inter-cloud space.
Such a partial cavity may be categorized by its completeness or shell measure with $\kappa<1$, depending on the fraction of the boundary from a perfect loop. 

ii) Shell (Sh $\kappa$): If the cloud's density is lower, the gas will be accumulated or snow-plowed around the shock front, making a shell structure.
The shell may be categorized by its completeness from $\kappa=1$ showing a perfect loop, or partial loops with $\kappa<1$.

iii) Partial/clumpy shell (Ps=Cs $\kappa$): Clouds are often more turbulent and clumpy. 
In such a case, the interaction front will produce more partial features such as a clumpy shell or an ensemble of partial arcs.
We categorize such a case by the fraction of the total partial arcs compared to a round loop by a factor of $\kappa$. 

\begin{figure}   
 	\begin{center}
\includegraphics[width=8cm]{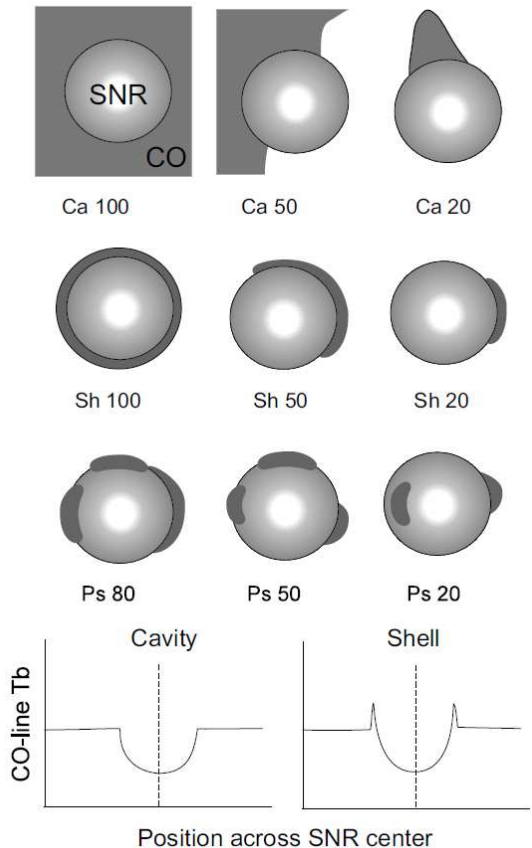} 
\end{center}
\caption{Shell types and shell measure .}
\label{shelltype}   
\end{figure}

\red{Figure \ref{shelltype} illustrates typical morphologies} of the CO brightness distribution around a SNR. 

\begin{figure*} 
\begin{center}    
\includegraphics[width=13cm]{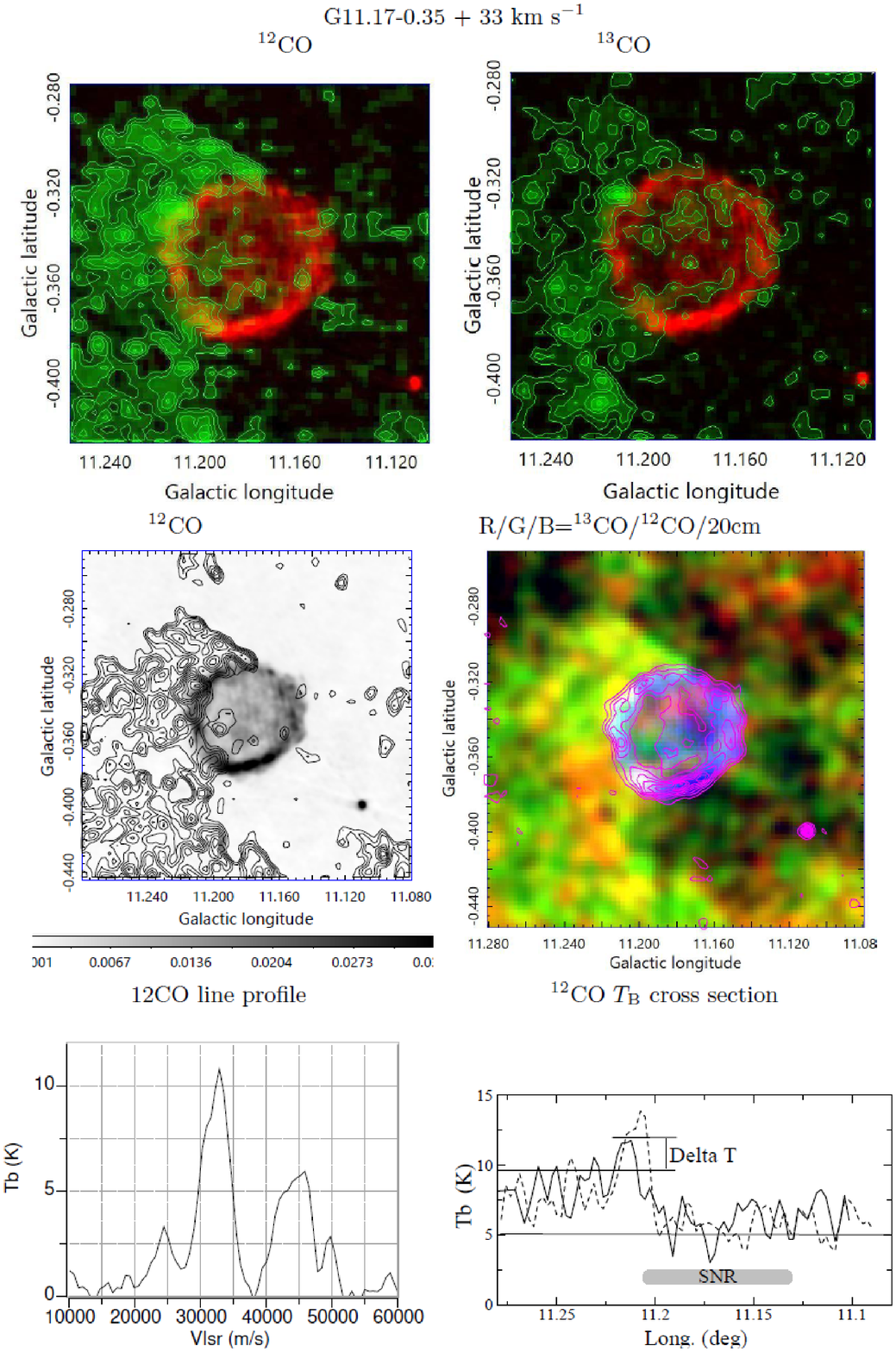}  
\end{center}
\caption{Example of molecular cavity/shell of Ca/Sh 50 toward SNR G11.17-0.35 at $\vlsr=+32.925$ \kms by (\red{top left}) $^{12}$CO contours from 7 K every 1 K on 20 cm in red; (\red{top right}) ibid, 7 K every 0.5 K, superposed on a grey-scale map of 20 cm radio continuum from 0 to 0.03 Jy beam$^{-1}$ ; (\red{middle left}) CO from 2 K every 1 K; (\red{middle right}) RGB color coded images of $^{13}$CO (red: auto), $^{12}$CO (green: auto) and 20 cm (blue: auto, magenta contour interval by 5 mJy beam$^{-1}$), (\red{bottom left}) $^{12}$CO line spectrum at the western edge; (\red{bottom right}) $^{12}$CO $\Tb$ across the SNR center along $b=-0.34 \deg$ (dash) and $-0.36 \deg$ (full line), showing the 'cavity' property. All figures of the studied SNRs are presented in the supplementary data described in the text.}   
\label{exG1118} 
\end{figure*}

\red{Descriptions of the properties} of the obtained maps are given in figure captions of individual objects in the atlas.
We here present an example for G11.17-0.35+32.975 \kms in figure \ref{exG1118}. 
This SNR is a typical bright shell in radio continuum, and its western half is apparently contacting, on the sky, with a half-cavity of a CO line cloud \citep{kilpa2016ApJ...816....1K}.
The RGB image indicates that the edge of the cavity facing the SNR does not show a signature of strong compression of gas, which would cause a red-color ($^{13}$CO) excess, if it existed.
\red{Also, the molecular mass (luminosity) along the edge of the cavity is far smaller than that expected by snow-plowed mass from inside the cavity. 
Such a structure may imply that the shell was created by dissociation of molecules, but not by a direct snow-plowing compression.}

\subsection{Distances}

\begin{figure}     
\begin{center}
\includegraphics[width=8cm]{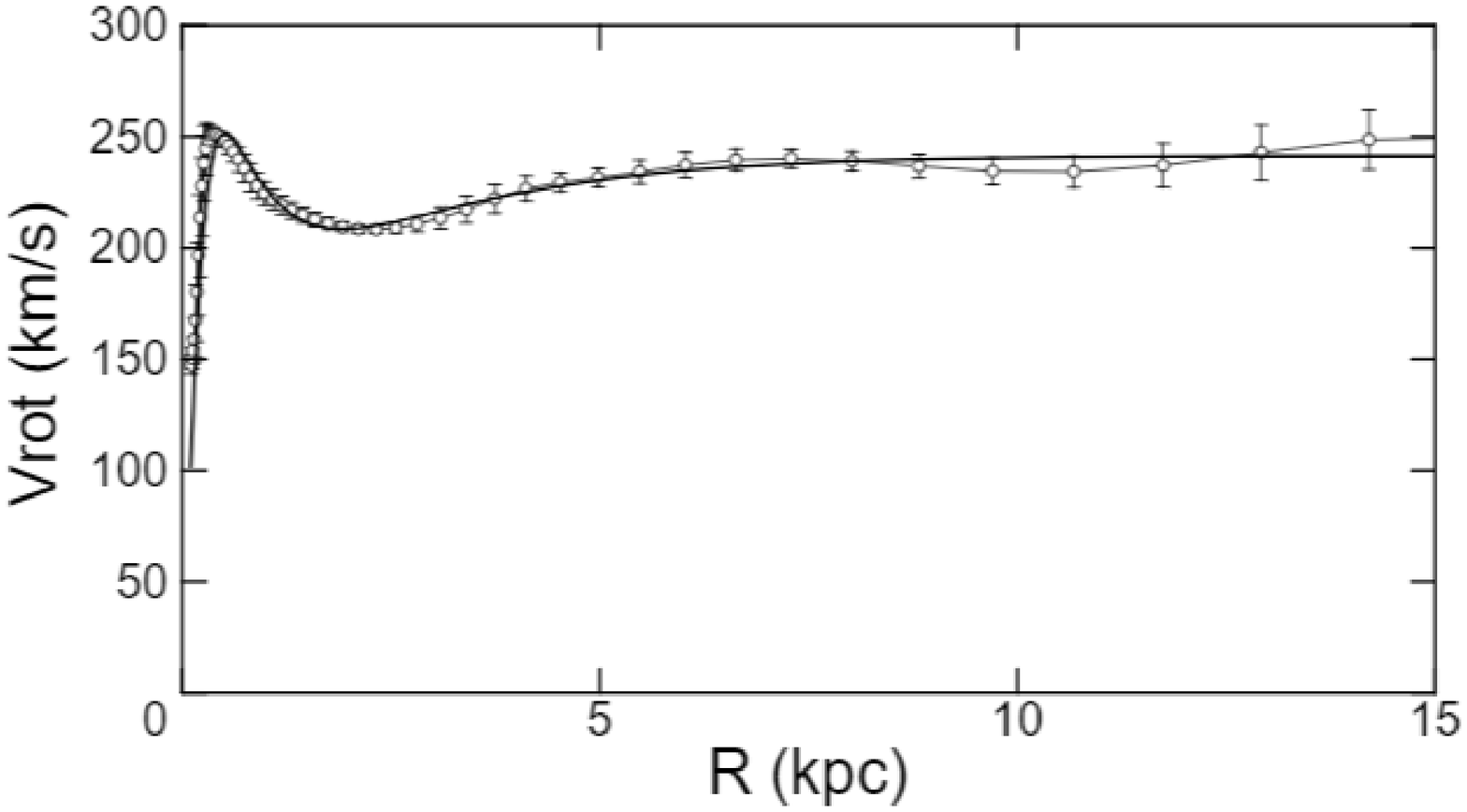} 
\end{center}
\caption{The most recent rotation curve (thin line with circles and standard errors) \citep{2020Galax...8...37S}, and the used model rotation curve expressed by equation \ref{eq-rc}. }
\label{rc}   
\end{figure}

The radial velocity, $v_r=\vlsr$, at a distance \red{$d$} orbiting around the Galactic Center is related to the circular rotation velocity $V(R)$ as a function of the galacto-centric distance $R$ as
\begin{equation}
v_r= \left({R_0 \over R} V(R) - V_0\right) \sin \ l,
\label{eq-vr}
\end{equation}
where $R$ is the galacto-centric distance related to \red{$d$} and galactic longitude $l$ by  
\begin{equation}
\red{d}=R_0 \cos \ l \pm \sqrt{R^2-R_0^2 {\rm sin}^2 l}
\label{eq-r}.
\end{equation}
We assume here $V_0=238$ \kms and $R_0=8.0$ \kms
\citep{honma2015PASJ...67...70H},
and adopt the most recent rotation curve derived by compilation of determined circular velocities in the last two decades
as shown in figure \ref{rc} 
\citep{2020Galax...8...37S}.
Here, we approximate the rotation curve by an analytic expression,
\begin{equation}
V(R)=\left( \frac{V_1}{[1+(R/a)^2]^{1/2}} + \frac{V_2}{1+(R/b)^2} \right) \left(\frac{R}{c}\right).
\label{eq-rc}
\end{equation}
The parameters, $V_1=67$, $V_2=1000$ \kms, $a=3.5$ kpc, $b=0.44$ kpc, $c=1$ kpc, were determined by iterative fitting of the function to the data by trial and error, until one gets satisfactory reproduction of the data within radius range, 1.4 to $\sim 10$ kpc, necessary for the present analysis.
The adopted curve is shown by the thick line in figure \ref{rc}. 

For a given set of $\vr$ and $l$, we can determine $R$ by iteration using equations \ref{eq-vr} and \ref{eq-rc}, and the distance \red{$d$} is obtained by equation \ref{eq-r}.
In table \ref{tab1} we list the determined distances and diameters of the SNR.
The errors are calculated using the uncertainty of radial velocity of the CO line in the measured value as well as the interstellar turbulence, $\delta \vlsr\sim 5$ \kms, and the uncertainty in the rotation velocity, $\delta \Vrot\sim 5$ \kms, propagating through the above equations to $d$.
The uncertainties in $R_0$ and $V_0$ are not included.
 
\subsection{Molecular mass} 

The molecular mass of associated clouds is one of the most essential quantities.
However, the present resolution, $20''\sim 0.5$ pc at 5 kpc for example, is a few orders of magnitudes wider than the expected thickness of shock-compressed filaments at the SNR fronts \citep{lucas+2020}.
So, we are not able to estimate meaningful mass of the directly associated molecular gas to the SNRs.

Instead, we here try to estimate the upper limit to the associated cloud for G11.17-0.35 as a typical example. 
Measuring the excess $\Tb$ at the edge of the SNR over that in the ambient emission outside SNR, the upper limit mass may be calculated by
\begin{equation}
    M\sim \mu\ \mH\ 2 \pi \kappa {R_{\rm s}}\ \delta R_{\rm beam} \Xco \int {\delta} T dv, 
\end{equation}
where $\mu\sim 2.6$ is the reduced mass per H$_2$ molecule for solar abundance, $\mH$ is the hydrogen mass, $\Xco \sim 2\times 10^{20}$ cm$^{-2}$ [K \kms]$^{-1}$ \citep{sofue+2020xco} is the conversion factor, $R_{\rm S}$ is the SNR radius, $\delta {R_{\rm beam}}=\theta_{\rm beam} r$ is the beam width at the object, $\kappa$ is the cavity/shell measure, and $\delta T$ is the excess brightness temperature of \co line at the intensity peak along the shell or the edge of cavity contacting the SNR.

For G11.17-0.35 at 33 \kms (figure \ref{exG1118}), we obtain $\delta T\sim 5$ K and $\kappa \sim 0.5$, and the possibly associated molecular mass is shown to be $M <\sim 10^2$ and $<\sim 10^3 \Msun$ for the near and far distances, respectively.  
Similar estimation applies to most of the observed partial CO shells in the analyzed SNR, but we do not present the results for individual objects, because the estimations are simply upper limits to the physically meaningful masses, which are supposed to be a few orders of magnitudes smaller, as discussed above.

\section{Summary}

We obtained a systematic search by morphology for cavity and/or shell structures of \co and \coth line emissions adjacent to 63 catalogued Galactic SNRs.
Such a search was possible only by careful inspection of individual channel maps of brightness temperature with high-velocity and high-angular resolutions from the FUGIN CO survey. 
The result is presented in the form of a table of kinematical distances of the CO shells, and an atlas of CO-line $\Tb$ maps as superposed on the radio continuum maps of the SNRs, which will be useful for general purpose for the investigation of the interaction between SNRs and ISM in the Galaxy.
 
\vskip 5mm
\noindent{bf Aacknowledgments}:\\
We are grateful to Prof. Masumichi Seta of Kwansei Gakuin University and Dr. Hidetoshi Sano of the National Astronomical Observatory of Japan for helpful advises, and to Mr. Yuya Tsuda of Meisei University for discussion.
The CO data were taken from the FUGIN CO survey obtained with the Nobeyama 45-m telescope, and retrieved from the JVO portal ({http://jvo.nao.ac.jp/portal}).
The data analysis was carried out at the Astronomy Data Center of the National Astronomical Observatory of Japan.
Radio continuum data were taken from the VGPS survey via the ATLASGAL data archives. 
The National Radio Astronomy Observatory is a facility of the National Science Foundation operated under cooperative agreement by Associated Universities, Inc.. 
This research is supported as part of the International Galactic Plane Survey through a Collaborative Research Opportunities grant from the Natural Sciences and Engineering Research Council of Canada.

Facilities: {Nobeyama 45-m, VLA, Effelsberg 100-m}

Software: {astropy \citep{2013A&A...558A..33A}}


\appendix
\section{SNR distribution}
\begin{figure*} 
\begin{center} 
\includegraphics[width=17cm]{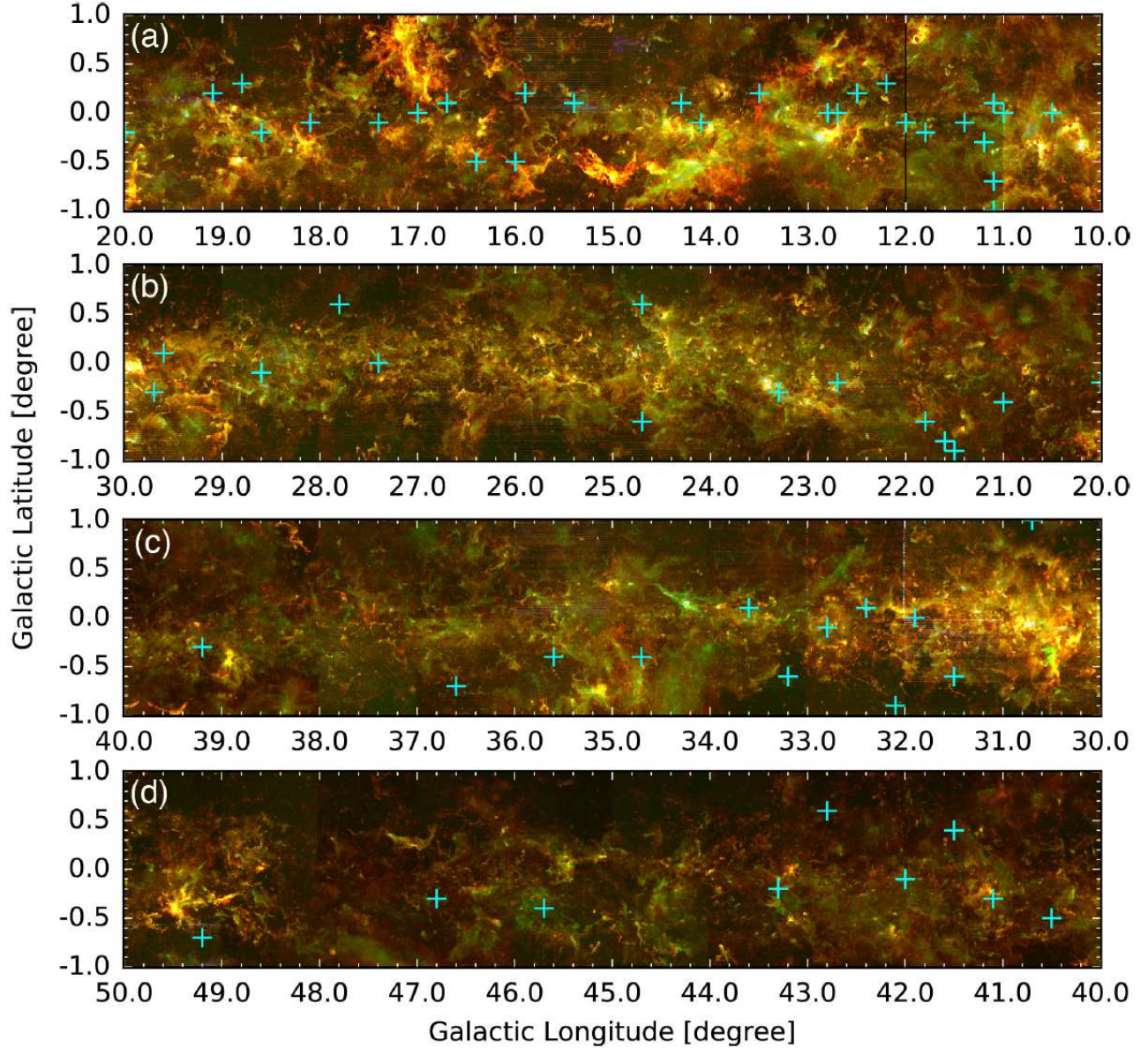} 
\end{center}
\caption{Green's SNRs (\red{cyan crosses}) superposed on the FUGIN CO map ({https://nro-fugin.github.io}) of the peak brightness temperatures of $^{12}$CO, $^{13}$CO and C$^{18}$O lines in red, green and blue, respectively \citep{2017PASJ...69...78U}. \red{The Galactic longitude range of (a) $10\deg \le l \le 20\deg$, (b) $20\deg \le l \le 30\deg$, (c) $30\deg \le l \le 40\deg$, and (d) $40\deg \le l \le 50\deg$, respectively.}}
\label{GreenOnFugin} 
 	\end{figure*}

Figure \ref{GreenOnFugin} shows the positions of the SNRs from the Green's catalogue on the CO line brightness map (red: $^{13}$CO, green: $^{12}$CO, blue: C$^{18}$O). 
  
\section{Table of SNRs} 

Table \ref{tab1} lists the analyzed objects and derived parameters for the candidate CO line features adjacent to the SNRs.
\begin{table*}[] 
\begin{center}
\caption{CO-line cavities and shells toward SNRs from the Green's catalogue.}
    \begin{tabular}{lllllllllllllll}
    \hline \hline  
(a) & (b) & (c) &(d) &(e)  & (f) & (g) & (h) & (i) & (j) & (k) &(l) & (m)\\
$l,b$ & $\vlsr$  &Size & SNR &$f_{\rm 1GHz}$  & Type$\dagger$& \red{$d_{\rm near}$} & \red{$d_{\rm far}$} & \red{$\delta d$} & $D_{\rm near}$ & $D_{\rm far}$&Name & \red{References} \\
($\deg,  \deg ~$)&{ (km/s)}  & $(' \times '$) & type &(Jy)  & $(\kappa$ in \%) & (kpc)  & (kpc) & (kpc) &(pc)& (pc)&&\\
    \hline 
11.00-0.05 &+40  & 11 9  & S & 1.3 & Sh 50&6&  11.6 & 0.3  & \red{12.0} & \red{33.4} & \\ 
11.1-0.7  &+33&  11 7&  & 1.0  &Ps 50&3.5 &  12.2 & 0.3 & 8.9 &31.2 &\\
(11.1 +0.1) &---  & 12 10  & S &2.3 &N\\ 
11.17-0.35& +33  & 4 4 & C &22 &Ca 50 &  3.7 & 12.0 & 0.3 & 4.3 & 14.0&&15\\
11.2+0.12 &+56  & 12 10 & S & 2.3 &Ps 30& 6 & 10.8&0.2& 11.6 & 25.7 \\
11.4-0.1 &+30  & 8 8  & S? &6  & Ca 60   & 3.4 &  12.3    &0.3    &8.0&   28.5 \\
---       &+50  &  & & &  Ca 50  &  4.1&11.5 &   0.3 &   9.6 &  26.9 \\  
11.89-0.23 & +49.8& 4 4 &F & 0.7 &Ca 50  &4.5&   11.1&    0.2&    5.3&   13.0\\
12.0 -0.1 &+37.4  & 7 7 ? &? &3.5 &Ps 50 &3.8&11.8  &  0.3    &7.7&   24.1\\ 
(12.2  +0.3)& --- &6 5&    S&       0.8  &N\\ 
(12.5  +0.2)& --- &6 5&    C?&      0.6  &N\\
(12.7  -0.0)& ---& 6&     S&       0.8 &N\\
(12.8  -0.0)& ---&    3    & C?    &  0.8 \\
13.45+0.14& +24  & 5 4  & S &3.5?   &Ca 50& 2.7 & 12.9 & 0.2 & 3.5 & 16.8 \\  
(14.1  -0.1)& ---& 6 5 &   S &  0.5 &N\\
(14.3  +0.1)&---&   5 4 &   S& 0.6 &N\\
15.42+0.16 &+34 & 15 14  & S &5.6  &Sh 60 &3.2 & 12.3 & 0.3 & 13.3 & 51.7  \\
15.9+0.2 &+29  & 7 5  & S?&5.0  & Ps 50  &2.8   &12.6    &0.3      \\
(16.0 -0.5) &--- & 15 10  & S &2.7 &N\\ 
(16.4 -0.5) &---  & 13 13  & S &4.6 &N \\ 
16.75+0.08& +47   & 4 4 & C &3.0  & Ca 90 &3.8 & 11.6 & 0.3 & 4.4 & 13.5\\ 
---       &  +62  &  &  & &  Ps 50  & 4.5 & 10.9 & 0.2 & 5.2 & 12.6\\
17.05-0.05& +31.0&  6 6& S& 0.4 & Ca 50 &2.8&   12.5&    0.3&    4.1&   18.2 &   \\
---& +93.4&   6 6& S&   1.4  &Ca 30&5.6&    9.7  &  0.2 &   8.2&   14.1   \\
18.1-0.1 &+49 & 8 8  & S &4.6  &Ps 50  & 3.7 & 11.5 & 0.3 & 8.7 & 26.7 &&10, 21\\
18.6  -0.2 & +66 &  6 6& S & 1.4 & Ca 50& 4.5& 10.7& 0.2& 7.8& 18.7 &&16\\
18.8+0.35 &+20& 17 11 & S &33  & Ps 60 & 1.9 &  13.3 & 0.4 & 7.4 & 52.8 &{Kes 67} &
9\\
(19.1 +0.2) &---  & 27 27  & S &10 &N\\ 
20.0-0.2 &+65   & 10 10 &F &10  & Ca 60& 4.3 & 10.7 & 0.2 & 12.6 & 31.2 &&11\\
(20.4 +0.1) &---& 8 8  & S? &9? &N\\ 
(21.0  -0.4)   &&  9 7 &   S    &   1.1 &N \\  
(21.5 -0.9) &--  & 5 5 & C &7  &No 20cm\\ 
 (21.6-0.8)   && 13  &   S    &   1.4 &No 20cm  \\  
21.8-0.6 &+83  & 20 20  & S &65    &Sh 70  & 5.0 & 9.9 & 0.2 & 28.9 & 57.6 &{ Kes 69} & 4, 21\\
22.7-0.2 &+75 & 26 26  & S? &33  & Ps 60 &4.6 & 10.2 & 0.2 & 34.7 & 76.9 &&14\\
23.3-0.3 &+70  & 27 27  & S &70 & Ps 70 &4.3 & 10.4 & 0.2 & 34.1 & 81.3&{ W41}&14\\
(23.6 +0.3) &---  & 10 10 &? &8? &N\\ 
24.7-0.6 &+60  & 15 15 &?  S? &8 & Ca 30& 3.8 & 10.7 & 0.3 & 16.7 & 46.7 &&19\\
24.7+0.6 &+112  & 30 15 & C? &20? &C/P 60& 6.3 & 8.2 & 0.4 & 39.1 & 50.6\\ 
\hline  
    \end{tabular} 
\end{center}
$^\dagger$ "N" stands for no possible cavity/shell in CO recognized. Columns (a),\red{(c),(d),and (e)} are from Green's catalogue\citep{green2009BASI...37...45G}.\\
Columns: (a) Galactic position; (b) CO line radial velocity from the present measurements using FUGIN; (c) apparent major and minor-axis sizes, $\theta_x,\theta_y$; (d) SNR type;   (e) radio flux at 1 GHz; 
(f) CO cavity or shell measure; (g) near solution of the distance for the CO radial velocity; (h) far distance; (i) distance error; (j)linear diameter for near distance $D=\sqrt{\theta_x \theta_y}d$; (k)  for far distance, (l) Name, 
\red{(m) References to other CO-line observations}:
1) \cite{1997ApJ...475..194K};
2) \cite{seta2004AJ....127.1098S};
3) \cite{2006ChJAA...6..210Y};
4) \cite{2009ApJ...691..516Z};
5) \cite{2009ApJ...694..376S};
6) \cite{2010ApJ...712.1147J};
7) \cite{2010AA...509L...4P};
8) \cite{2011ApJ...727...43S};
9) \cite{2012AA...547A..60P};
10) \cite{2013MNRAS.433.1619P};
11) \cite{2013AA...554A..73P};
12) \cite{2013ApJ...768..179Y};
13) \cite{2014ApJ...793...95Z};
14) \cite{2015ApJ...811..134S};
15) \cite{kilpa2016ApJ...816....1K};
16) \cite{2016MNRAS.458.2813V};
17) \cite{2017ApJ...836..211S};
18) \cite{2017ApJ...843..119R};
19) \cite{rana2018MNRAS.477.2243R}; 
20) \cite{2018ApJ...864..161K}; 
21) \cite{lee+2020arXiv201107711L}. 
    \label{tab1}
\end{table*}  

\setcounter{table}{1}
\begin{table*}[] 
\begin{center}
\caption{Continued.}
    \begin{tabular}{lllllllllllllllll}
    \hline \hline   
$l,b$ & $\vlsr$  &Size & SNR &$f_{\rm 1GHz}$ & Type & \red{$d_{\rm near}$} & \red{$d_{\rm far}$} & \red{$\delta d$} & $D_{\rm near}$ & $D_{\rm far}$ & Name & \red{References}  \\
($\deg,  \deg ~$)&{ (km/s)}  & $(' \times '$) & type &(Jy)  & $(\kappa$ in \%) & (kpc)  & (kpc) &(kpc) &(pc)&(pc)  \\
    \hline   
27.4+0.0 &+101  & 4 4  & S &6  &Ca 60 &5.8 & 8.4 & 0.4 & 6.8 & 9.8&{4C-04.7}&15\\
(27.8 +0.6) &---  & 50 30 &F &30 &N\\ 
28.62-0.10& +86  & 13 9  & S &3?  & Ca 60  & 5.0 &   9.0    &0.3 &  15.7 &  28.5 &&19\\

29.6  +0.1&+99.2&  5  &   S    &   1.5? &Sh 50 &5.9    &8.0&    0.5&    8.6&   11.7& &15\\
  
29.70-0.26 &+52  & 3 3 & C &10 &Ca 50& 3.3 & 10.6 & 0.3 & 2.9 & 9.3& { Kes 75} &5\\
---        &+112  &  &  & & Ps 40  &3.3 & 10.6 & 0.3 & 2.9 & 9.3\\
31.5-0.6 & +87.5 & 18 18?  & S? &2?  & Ps 50 & 5.2 & 8.4 &0.3 & 27 &44&\\ 
----- & +97 &   &  & &  Ca 40& 6.0 & 7.6 & 0.5 & 31 & 40 & \\ 
31.9+0.0 &+107   & 7 5  & S &25  &Ca 50& 6.8 & 6.8 & -- & 12 & 12 &  3C391&18\\
32.1-0.9 &+95  &  40 40? & C? & ?  & N & 5.9& 7.7 &0.5 & 69 &89&\\ 
32.4+0.1 & +10.8 & 6 6  & S & 0.25?  & Ca 100& 0.79 & 12.7 & 0.2 & 1.4 & 22 &\\ 
----- & +42.6 &   &  &  &Ca 30& 2.7 & 10.8 & 0.2 &4.8&18.8&&15\\
32.8 -0.1 & +74 & 17 17  & S? &11?  &Ps 10 & 4.4 & 9.0 & 0.25 &22 & 45 & Kes 78\\
----- & +103 &   &  & &  Ca 30& 6.7 & 6.7 &-- & 33 & 33 & Kes 78\\ 
33.2 -0.6 & +54  &18  18  & S &3.5  & Ps 20& 3.3 & 10.1 & 0.2 & 17 & 53\\
----- & +91 &   &  & &  Ca 10 & 5.7 & 7.7 & 0.5 & 30 & 40 &\\
33.7+0.05 &+70  & 10 10  & S &20  & Ca 70& 4.2 & 9.1 & 0.3 & 12.3 & 26.4 & { Kes 79} &20\\
34.7-0.4 &+40  & 35 27 & C &250 & Ca 60&  2.6 & 11 & 0.2 & 23 & 95 &{ W44} &2\\
----- &+52  &  &  & &  Ca 75& 3.2 & 9.9 & 0.2 & 29 & 89 &{ W44} & 12\\
35.6 -0.4 &+55  & 15 11  & S? &9 & Ca 40 &3.4& 9.6& 0.2&13&36& &7\\
----- & +90 &   &  & &  Pa 20& 6.5 & 6.5 & -- & 24 & 24 &\\
36.6-0.7 & +57 & 25 25?  & S? & 1.0 & Ca 20 & 3.5 & 9.3 & 0.3 & 26 & 68 &\\
----- & +79 &   &  & & Ca 20 & 5.1 & 7.8 & 0.4 & 37 & 57 &\\
39.2-0.3& +51  & 8 6 & C & 18  & Ca 30&  3.2 & 9.2 & 0.3 & 6.5 & 18.5& {3C396} & 21\\
---        & +67  &  &  &  &   Ca 60& 4.3 & 8.1 & 0.5 & 8.7 & 16.2 &&8\\
40.5 -0.5 &+58 & 22 22  & S &11 & Ca 70 &3.8& 8.4 &0.3 &24 & 54 &&3\\
41.1 -0.3 & +32 & 4.5 2.5  & S &25 & Ca 60 &2.1 &10.0 & 0.3 &2.0 & 9.7 &3C397 &6\\
----- & +38 &   &  & &  Ca100 & 2.5 & 9.6 & 0.3 & 2.4 & 9.4 &\\
41.5+0.4 & +58 &10 10& S? & 1?  & Ca 50 &3.8 & 8.2 & 0.4 & 11 & 24  &\\ 
42.0-0.1 & +66 & 8x8  &  S? & 0.5?  & Ca 60 & 4.6 & 7.3 & 0.5 & 11 & 17&\\ 
(42.8 +0.6) &  & 24 24  & S &3?  & N\\
43.3-0.2 &+10 & 4 3  & S &38  & Ca 50&  0.7 & 11 & 0.3 & 0.7 & 11 &{ W49B} & 21\\
----- &+45 &  &  & &  Ca 60&  3.0 & 8.7 & 0.3 & 3.0 & 8.7 &{ W49B}&13, 21\\
----- &+62 &  &  & &  Ca 100&  4.4 & 7.3 & 0.7 & 4.4 & 7.3&{ W49B} & 21\\
45.7 -0.4 & +26  & 22 22  & S &4.2?& Pa 40 & 1.8 & 9.4 & 0.3 & 11& 60\\ 
----- & +48.5 &   &  & &  Pa 20 & 3.4 & 7.8 & 0.4 &22 & 50 &\\
46.8-0.3 &+52  & 15  & S &17  & Ca 70 &  3.9 & 7.1 & 0.5 & 17 & 31 &{ HC30}\\
49.2 -0.7 & +50 & 30 30  & S? &160?  &Pa 30 & 4.1 & 6.4 & 0.7 & 35 & 56 &{ W51C} &1\\
----- & +60 & 30 30  & S? &160?  & Ca 50 & 5.2 & 5.2 & -- & 46 & 46 &{W51C}\\ 
\hline   
\hline
 205.5  +0.5 & +10 & 220 &    S &    140  & N & 0.98 & --- & 0.3& 63 & --- &  Monoceros & 17\\
----- & +20 &  & &   &    N & 2.2 & --- & 0.3& 139 & --- &  Monoceros\\
 213.0  -0.6& +9 & 160x140?& S&21  &  0.4  N &0.7 & --- & 0.3& 32 & --- & &17\\
----- & +21 & & &  &   N &1.8 & --- & 0.3& 80 & --- &\\
\hline  
    \end{tabular}   
\end{center}
    \label{tablecontinued1}
\end{table*} 

\section{Figures}

\begin{figure*} 
\begin{center}        
\includegraphics[width=14cm]{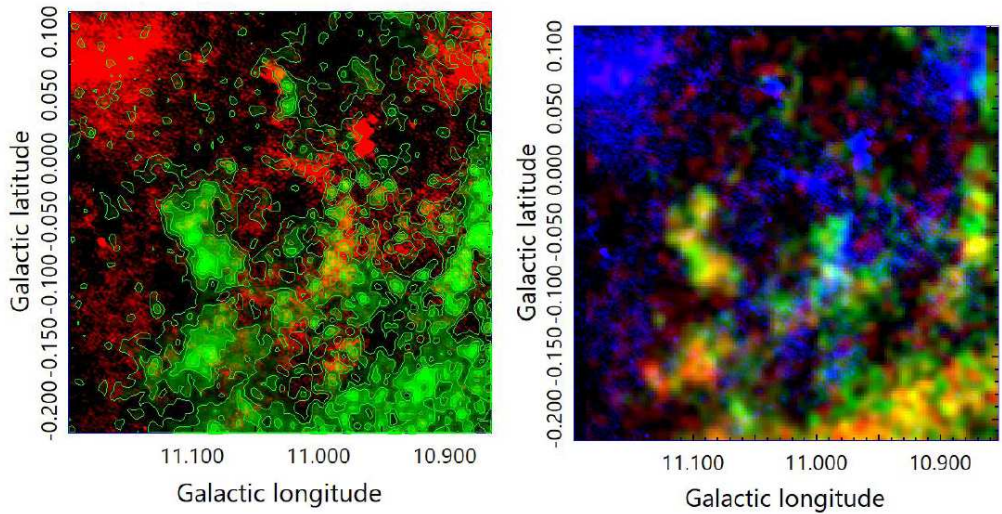}  
\end{center}
\caption{Same as figure \ref{exG1118}, but for SNR G11.00-0.05+40.725, 40.075 km s$^{-1}$: 
(Left) $^{12}$CO contours superposed on the radio continuum map in red. Contours start at 2K by step 1K.
(Right) Two color composite image of CO $\Tb$, red and green showing $^{12}$CO and $^{13}$CO respectively, superposed on 20-cm radio map. 
}   
\label{G1100} 
 	\end{figure*}

Figures \ref{G1100} to 51 show the analyzed results on individual SNRs. 

\noindent{\bf 
Full atlas including all figures from 5 to 51 is available in this URL: 
https://nro-fugin.github.io/2020-apjs-CO-Shell-Atlas-SNR-FUGIN-IX.pdf
}

 	\end{document}